\documentclass[a4paper,11pt]{article}
\pdfoutput=1 

\usepackage{jcappub} 

\usepackage{bbm}
\usepackage{cleveref}
\usepackage{mathtools}
\usepackage{upgreek}

\crefname{equation}{}{}
\Crefname{equation}{}{}
\crefname{figure}{figure}{figures}
\Crefname{figure}{Figure}{Figures}
\crefname{table}{table}{tables}
\Crefname{table}{Table}{Tables}
\crefname{section}{section}{sections}
\Crefname{section}{Section}{Sections}

\newcommand{\dif}{\mathrm{d}}
\newcommand{\ex}{\mathrm{e}}
\newcommand{\im}{\mathrm{i}}

\newcommand{\ba}{\mathrm{b}}
\newcommand{\dm}{\mathrm{d}}
\newcommand{\rad}{\mathrm{r}}
\newcommand{\bary}{\mathrm{B}}
\newcommand{\eq}{\mathrm{eq}}

\title{Baryon-photon interactions in Resummed Kinetic Field Theory}

\author[a,1]{Ivan Kostyuk,\note{Corresponding author.}}
\author[b]{Robert Lilow,}
\author[c]{and Matthias Bartelmann}

\affiliation[a]{Max Planck Institute for Astrophysics, 85748 Garching, Germany}
\affiliation[b]{Department of Physics, Technion, Haifa 3200003, Israel}
\affiliation[c]{Institut f\"ur Theoretische Physik, Heidelberg University, Philosophenweg 16, 69120 Heidelberg, Germany}

\emailAdd{ivkos@mpa-garching.mpg.de}
\emailAdd{rlilow@campus.technion.ac.il}
\emailAdd{bartelmann@uni-heidelberg.de}

\abstract{
We explore how interactions between baryons and photons can be incorporated into Kinetic Field Theory (KFT), a description of cosmic structure formation based on classical Hamiltonian particle dynamics. In KFT, baryons are described as effective mesoscopic particles which represent fluid elements governed by the hydrodynamic equations. In this paper, we modify the mesoscopic particle model to include pressure effects exerted on baryonic matter through interactions with photons. As a proof of concept, we use this extended mesoscopic model to describe the tightly coupled baryon-photon fluid between matter-radiation equality and recombination. We show that this model can qualitatively reproduce the formation of baryon-acoustic oscillations in the cosmological power spectrum.
}

\arxivnumber{2302.?????}


\begin{document}
\maketitle
\flushbottom

\section{Introduction}
One of the main interests in cosmology is to gain an understanding of how structures formed in the Universe from small Gaussian perturbations at the time of the CMB decoupling to the highly non-linear structures we see today. While the linear growth of structures is well understood, it becomes far more difficult to describe structure growth once it enters the non-linear regime.

One approach to address this problem are cosmological large-scale simulations such as IllustrisTNG \cite{springel_first_2018} and EAGLE \cite{schaye_eagle_2015}. These simulations are able to reliably capture structure formation far into the non-linear regime. However, they are computationally extremely expensive, which limits possible studies of the vast landscape of potential cosmological models.

To address this problem, as well as to gain a deeper insight into the underlying physical processes shaping the formation of non-linear structures, perturbative analytic approaches were developed. Examples hereof can be found in \cite{crocce_renormalized_2006,matarrese_resumming_2007,matsubara_resumming_2008,pietroni_coarse-grained_2012,blas_cosmological_2014,porto_lagrangian-space_2014,senatore_ir-resummed_2015} and a comprehensive review in \cite{bernardeau_large-scale_2002}. However, one problem these theories often face is their treatment of shell-crossing. This is the process in which multiple streams of dark matter flow through each other, such that there are points in space with a multi-valued velocity field. Recently, perturbative approaches started being developed to address the problem of shell-crossing \citep[e.g.][]{mcdonald_large-scale_2018,pietroni_structure_2018}.

The approach employed in this work, Kinetic Field Theory (KFT) \citep[e.g.][]{bartelmann_microscopic_2016,bartelmann_cosmic_2019}, is based on a path integral description of classical Hamiltonian particle dynamics \citep{gozzi_hidden_1988,gozzi_hidden_1989,das_field_2012,das_newtonian_2013}. The main advantage of this new approach is that the theory captures the full Hamiltonian evolution of the particles in phase space. There is thus no need to construct a smooth, single-valued velocity field, as is the case in the standard Eulerian and Lagrangian perturbation theories \citep{bernardeau_large-scale_2002}. Instead, the multiple velocity field values during stream-crossing are taken into account. A detailed comparison between the Eulerian perturbative theory and KFT can be found in \cite{kozlikin_first_2021}.

But even the most accurate treatment of dark matter dynamics yields an incomplete description of structure formation. In the late Universe, baryonic effects have a significant impact on structure formation on scales $\lesssim 1 \, h^{-1}\,\mathrm{Mpc}$. In the early Universe, they are essential for the formation of the Baryon Acoustic Oscillations (BAOs) \citep{sunyaev_small-scale_1970,peebles_primeval_1970}. Recently, in \cite{geiss_resummed_2021} KFT was therefore further extended to describe a two-particle type system composed of both dark and baryonic matter. The frequent non-gravitational interactions of individual baryonic particles render a perturbative treatment of their microscopic dynamics in KFT infeasible. Therefore, we instead describe baryonic matter through effective mesoscopic particles \cite{viermann_model_2018,geiss_resummed_2019}, representing fluid elements that follow the hydrodynamic equations. A mesoscopic particle thus has to be large enough to establish a local thermal equilibrium inside of it. Compared to our cosmological scales of interest, however, a mesoscopic particle can still be treated as point-like. In addition, it was shown in \cite{geiss_resummed_2019}, that a consistent perturbative description of mesoscopic baryons in KFT requires to work in the Resummed KFT (RKFT) framework developed in \cite{lilow_resummed_2019}. While standard KFT perturbation theory expands in orders of the interaction potential, RKFT expands in orders of (phase space) density correlation functions. The main advantage of this new expansion scheme is that it leads to a resummation of infinite subsets of the standard KFT perturbative expansion.

Most of the baryonic effects on structure formation are not a result of baryonic gas dynamics on their own, but caused by the interactions of baryons and photons. This includes radiative cooling and heating effects in the late Universe as well as the strong coupling between baryons and photons before recombination. The aim of this paper is to explore the description of baryon-photon interactions in KFT. Since photons do not follow classical particle dynamics, they cannot be treated explicitly as a third particle type. Instead, we further extend the mesoscopic particle model to also incorporate an effective treatment of the interactions between photons and baryons.

In this work, we specifically study the tight-coupling regime before recombination, i.e.~at $z \gtrsim 1100$. In this regime, the mean free path of photons stays significantly below the scale of the mesoscopic particles. Hence, we can directly associate the mesoscopic particles with fluid elements of the coupled baryon-photon fluid. Furthermore, this epoch of structure formation is well-described by linear gravitational dynamics, allowing us to investigate the formation of BAOs in the matter power spectrum in the lowest perturbative order of RKFT. The goal of this proof of concept application is to lay the groundwork for a more general incorporation of radiative effects into the KFT formalism. The specific treatment of radiative effects on small nonlinear scales in the late Universe will be the subject of future work.

This paper is structured as follows. In \cref{sec:theory} we summarize the (R)KFT formalism for two particle types. In \cref{sec:application} we demonstrate its application to the tightly coupled baryon-photon fluid in the early Universe. We present the resulting linear power spectra and show the emergence of BAOs. Finally, we discuss our results and the next steps of development in \cref{discussion}.

\section{KFT with two particle types}
\label{sec:theory}
To describe the interactions between dark matter and a baryon-photon fluid, we employ the two-particle-type formulation of (R)KFT presented in \cite{geiss_resummed_2021}. The following section summarizes the main aspects of this formalism necessary to understand its subsequent application to structure formation before recombination. We refer readers interested in more details of this formalism and its derivation to \cite{geiss_resummed_2021} as well as the underlying work in \cite{bartelmann_microscopic_2016,fabis_kinetic_2018,lilow_resummed_2019}.

\subsection{Generating functional}
In the KFT formalism, the classical dynamics of interacting particles in phase space is encoded in a generating functional that can then be used to calculate any macroscopic moments of interest, e.g.~density correlation functions.

We denote the phase-space coordinates of an individual particle of species $\alpha \in \{\ba, \dm \}$ (corresponding to the baryon-photon fluid and dark matter) by $\vec{x}^{\alpha}_i \coloneqq (\vec{q}^\alpha_i,\vec{p}^\alpha_i)$. The index $i = 1, \dotsc, N^\alpha$ labels a specific particle with spatial and momentum coordinates $\vec{q}^\alpha_i$ and $\vec{p}^\alpha_i$, respectively. The coordinates of all $N^\alpha$ particles of species $\alpha$ are collected in
the tensor
\begin{equation}
	\boldsymbol{x}^\alpha \coloneqq \sum_{i=1}^{N^\alpha} \, \vec{x}^\alpha_i \otimes \vec{e}^\alpha_i \,,
\end{equation}
where $\{\vec{e}^\alpha_i \}$ denotes the canonical basis vectors with components $(\vec{e}^\alpha_i)_j = \delta_{ij}$, and no summation over $\alpha$ is implied. The state of the whole system is then summarized in the tensor tuple $\boldsymbol{\vec{x}}\coloneqq(\boldsymbol{x}^\mathrm{b},\boldsymbol{x}^\mathrm{d})$. We define the scalar product between two general tensor tuples $\boldsymbol{\vec{a}}$ and $\boldsymbol{\vec{b}}$ as
\begin{equation}
	\boldsymbol{\vec{a}}\cdot \boldsymbol{\vec{b}} \coloneqq \boldsymbol{a}^\mathrm{b}\cdot\boldsymbol{b}^\mathrm{b} + \boldsymbol{a}^\mathrm{d}\cdot\boldsymbol{b}^\mathrm{d} = \sum_{i=1}^{N^\ba} \, \vec{a}^\mathrm{b}_i \cdot \vec{b}^\mathrm{b}_i + \sum_{i=1}^{N^\dm} \, \vec{a}^\mathrm{d}_j \cdot \vec{b}^\mathrm{d}_j.
\end{equation}.

There are two ingredients characterizing the system that enter the generating functional: (i) The initial phase-space probability distribution, $P\bigl(\vec{\boldsymbol{x}}^{(\mathrm{i})}\bigr)$, describes our (generally only probabilistic) knowledge of the particles' phase-space coordinates at a given initial time $t_\im$. (ii) The equations of motion, $\vec{\boldsymbol{E}}[\vec{\boldsymbol{x}}] = 0$, describe how these coordinates subsequently evolve. The generating functional is then given by
\begin{equation}
    Z[\vec{\boldsymbol{J}},\vec{\boldsymbol{K}}] \coloneqq \! \int \! \mathrm{d}\vec{\boldsymbol{x}}^{(\im)} P\bigl(\vec{\boldsymbol{x}}^{(\im)}\bigr) \! \int\limits_{\vec{\boldsymbol{x}}^{(\im)}} \!\! \mathcal{D}\vec{\boldsymbol{x}} \! \int \! \mathcal{D}\vec{\boldsymbol{\chi}} \, \exp \bigg\{ \im \!\! \int \!\! \mathrm{d}t \bigg( \vec{\boldsymbol{\chi}} \cdot \vec{\boldsymbol{E}}[\vec{\boldsymbol{x}}] + \vec{\boldsymbol{\chi}} \cdot \vec{\boldsymbol{K}} + \vec{\boldsymbol{x}} \cdot \vec{\boldsymbol{J}} \bigg) \bigg\} \,.
    \label{generating func}
\end{equation}
It sums over all possible trajectories of the particles in phase space averaged over their initial probability distribution. The integral of the first term in the exponential over the auxiliary field $\vec{\boldsymbol{\chi}}$ is a convenient mathematical representation of the Dirac delta distribution $\updelta_\textsc{d}\bigl[\vec{\boldsymbol{E}}[\vec{\boldsymbol{x}}]\bigr]$, which ensures that only trajectories satisfying the equations of motion contribute. In addition, the source fields $\vec{\boldsymbol{J}}(t)$ and $\vec{\boldsymbol{K}}(t)$ have been introduced to obtain moments of the fields $\vec{\boldsymbol{x}}$ and $\vec{\boldsymbol{\chi}}$, respectively, by applying suitable functional derivatives,
\begin{equation}
	\left\langle \vec{\boldsymbol{x}}(t) \otimes \cdots \otimes \vec{\boldsymbol{\chi}}(t')\right\rangle = \frac{\updelta}{\im \updelta \vec{\boldsymbol{J}}(t)} \otimes \cdots \otimes \frac{\updelta}{\im \updelta \vec{\boldsymbol{K}}(t')} \, Z[\vec{\boldsymbol{J}},\vec{\boldsymbol{K}}] \bigg|_{\vec{\boldsymbol{J}},\vec{\boldsymbol{K}}=0} .
	\label{moments}
\end{equation}

The equations of motion for a particle of species $\alpha$ are
\begin{equation}
	\boldsymbol{E}^\alpha[\boldsymbol{x}] = (\partial_t + \boldsymbol{F}) \, \boldsymbol{x}^\alpha +  \bigl(0, \boldsymbol{\nabla}_{q^\alpha}\bigr) V^\alpha = 0 \,,
	\label{EOM}
\end{equation}
where $\boldsymbol{F}$ encapsulates the linear part of the equations of motion, while the non-linear contribution is given by the gradient of the potential $V^\alpha$. The potential experienced by a particle of a given species consists of the individual contributions generated by each particle of both species,
\begin{equation}
	V^\alpha(\vec{q},t) = \sum_{j=1}^{N^\mathrm{b}} v^{\alpha \mathrm{b}}(\vec{q}-\vec{q}^{\,\ba}_j,t) + \sum_{j=1}^{N^\mathrm{d}}
	v^{\alpha \mathrm{d}}(\vec{q}-\vec{q}^{\,\dm}_j,t) \,.
	\label{eq:total_interaction_potential}
\end{equation}
Here, $v^{\alpha \gamma}(\vec{q}-\vec{q}^{\,\gamma}_j,t)$ is the potential contribution generated by the $j$-th particle of species $\gamma$ and experienced by a particle of species $\alpha$ at position $\vec{q}$.

The initial phase-space probability distribution $P(\boldsymbol{\vec{x}}^\mathrm{(i)})$ of the particles is obtained through Poisson sampling of the initial density and momentum distribution. For the early initial times considered here, the initial density contrast and momentum fields follow zero-mean Gaussian distributions, fully defined by their covariance. The derivation proceeds analogous to the one-particle-type case in \cite{bartelmann_microscopic_2016,fabis_kinetic_2018}, yielding
\begin{equation}
    P\bigl(\vec{\boldsymbol{x}}^{(\im)}\bigr) = \frac{V^{-(N^\ba+N^\dm)}}{\sqrt{(2\pi)^{3(N^\ba+N^\dm)} \det \boldsymbol{C}_{pp}}} \, \hat{\mathcal{C}} \bigg( \frac{\partial}{\im \partial \vec{\boldsymbol{p}}^{(\im)}} \bigg) \, \exp \bigg\{ -\frac{1}{2} \vec{\boldsymbol{p}}^{(\im)\intercal} \, \boldsymbol{C}^{-1}_{pp} \, \vec{\boldsymbol{p}}^{(\im)}  \bigg\} \,.
    \label{MPH: Initial Distribution}
\end{equation}
Here, $V$ is the volume containing the $N^\ba + N^\dm$ particles, $\boldsymbol{C}_{pp}$ denotes the initial momentum covariance matrix, and $\hat{\mathcal{C}}$ is a polynomial derivative operator that depends on the initial density contrast covariance matrix $\boldsymbol{C}_{\delta\delta}$ as well as the initial cross covariance matrix between density contrast and momentum $\boldsymbol{C}_{\delta p}$. The precise definition of $\hat{\mathcal{C}}$ can be found in \cite{fabis_kinetic_2018}. The initial density contrast and momentum fields are statistically homogeneous and isotropic. In addition, the initial momentum field can safely be assumed to be irrotational, allowing us to express it in terms of the negative initial momentum divergence $\theta^{(\im)} \coloneqq - \vec{\nabla} \cdot \vec{p}^{(\im)}$. Then, the different covariance matrix elements are fully determined by the initial auto- and cross-power spectra of the density contrast and the negative momentum divergence of the different particle species, $P^{\alpha\gamma\,\mathrm{(i)}}_{\delta\delta}$, $P^{\alpha\gamma\,\mathrm{(i)}}_{\delta\theta}$ and $P^{\alpha\gamma\,\mathrm{(i)}}_{\theta\theta}$.

\subsection{Collective fields}
When investigating cosmic structure formation with KFT, we are interested in the collective behaviour of the cosmic density field rather than individual particles. To extract this information, we introduce the collective number density field $\vec{\Phi}_n = (\Phi_n^\ba, \Phi_n^\dm)$, the components of which are
\begin{equation}
    \Phi_n^\alpha(\vec{q},t) = \sum_{j=1}^{N^\alpha} \updelta_\textsc{d}\bigl(\vec{q}-\vec{q}_j^\alpha(t)\bigr) \,.
\end{equation}
In addition, it is convenient to introduce the collective response field $\vec{\Phi}_B = (\Phi^\ba_B, \Phi^\dm_B)$,
\begin{equation}
    \Phi^\alpha_B(\vec{q},t) \coloneqq \sum_{j=1}^{N^\alpha} \vec{\chi}^\alpha_{p_j}(t) \cdot \vec{\nabla}_q \, \updelta_\textsc{d}\bigl(\vec{q}-\vec{q}_j^\alpha(t)\bigr) \,,
\end{equation}
which describes how the particle momenta are changed by a given interaction potential.
In the following, it is preferable to represent the collective fields in Fourier space,
\begin{align}
    \Phi^\alpha_n(\vec{k},t) &= \sum_{j=1}^{N^\alpha} \ex^{\im \vec{k}\cdot\vec{q}^\alpha_j(t)} \,,
    \label{density collective} \\
    \Phi^\alpha_B(\vec{k},t) &= \sum_{j=1}^{N^\alpha} \im \vec{k} \cdot\vec{\chi}^\alpha_{p_j}(t) \, \ex^{-\im \vec{k}\cdot\vec{q}^\alpha_j(t)} \,.
    \label{response field collective}
\end{align}
We further define the corresponding collective field operators $\hat{\Phi}^\alpha_n$ and $\hat{\Phi}^\alpha_B$ by replacing all occurrences of the fields $\vec{\boldsymbol{x}}$ and $\vec{\boldsymbol{\chi}}$ by functional derivatives with respect to the matching source fields  $\vec{\boldsymbol{J}}$ and $\vec{\boldsymbol{K}}$. Then, collective-field cumulants, i.e.~connected correlation functions of $\Phi^\alpha_n$ and $\Phi^\alpha_B$, can be obtained by acting with these operators on the logarithm of the generating functional,
\begin{align}
	G^{\alpha_1 \dots \alpha_{l_n} \, \gamma_1 \dots \gamma_{l_B}}_{n \cdots n \, B \cdots B}(1,\dots,l_n,1',\dots,l_B') &= \left\langle \prod_{u=1}^{l_n} \Bigl(\Phi^{\alpha_u}_n(u)\Bigr) \, \prod_{r=1}^{l_B} \Bigl(\Phi^{\gamma_r}_{B}(r')\Bigr) \right\rangle_c
	\label{Interacting collective-field cumulants} \\ 
	&= \prod_{u=1}^{l_n} \Bigl(\hat{\Phi}^{\alpha_u}_n(u)\Bigr) \, \prod_{r=1}^{l_B} \Bigl(\hat{\Phi}^{\gamma_r}_{B}(r')\Bigr) \, \ln Z[\vec{\boldsymbol{J}},\vec{\boldsymbol{K}}]\bigg|_{\vec{\boldsymbol{J}},\vec{\boldsymbol{K}}=0} \,. \nonumber
\end{align}
Here, we introduced the shorthand notation $(\pm m) \coloneqq (\pm\vec{k}_m, t_m)$ to combine Fourier-space and time arguments.

To perturbatively compute these cumulants, the generating functional \cref{generating func} can be split into a free and an interacting part,
\begin{align}
    Z[\vec{\boldsymbol{J}},\vec{\boldsymbol{K}}] &= \! \int \! \mathrm{d}\vec{\boldsymbol{x}}^{(\im)} P\bigl(\vec{\boldsymbol{x}}^{(\im)}\bigr) \! \int\limits_{\vec{\boldsymbol{x}}^{(\im)}} \!\! \mathcal{D}\vec{\boldsymbol{x}} \! \int \! \mathcal{D}\vec{\boldsymbol{\chi}} \, \exp \bigg\{ \im S_\mathrm{I}[\vec{\boldsymbol{x}},\vec{\boldsymbol{\chi}}] + \im \!\! \int \!\! \mathrm{d}t \bigg( \vec{\boldsymbol{\chi}} \cdot \bigl[(\partial_t + \boldsymbol{F}) \, \vec{\boldsymbol{x}} + \vec{\boldsymbol{K}}\bigr] + \vec{\boldsymbol{x}} \cdot \vec{\boldsymbol{J}} \bigg) \bigg\} \nonumber \\
    &= \ex^{\im \hat{S}_\im} \int \! \mathrm{d}\vec{\boldsymbol{x}}^{(\im)} P\bigl(\vec{\boldsymbol{x}}^{(\im)}\bigr) \, \ex^{\im \! \int \! \mathrm{d}t \, \vec{\boldsymbol{J}} \cdot \vec{\boldsymbol{x}}^\mathrm{lin}} \eqqcolon \ex^{\im \hat{S}_\im } \, Z_0[\vec{\boldsymbol{J}},\vec{\boldsymbol{K}}] \,.
    \label{free and interacting}
\end{align}
Here, $\vec{\bf{x}}^\mathrm{lin}$ corresponds to the free trajectory, i.e.~the solution to the linear part of the equations of motion,
\begin{equation}
	\boldsymbol{x}^{\mathrm{lin},\alpha}(t) \coloneqq \boldsymbol{\mathcal{G}}^{\mathrm{R}\alpha}(t,t_\im) \, \boldsymbol{x}^{\alpha(\im)} - \int_{t_\im}^\infty \mathrm{d}t'\ \boldsymbol{\mathcal{G}}^{\mathrm{R}\alpha}(t,t') \, \boldsymbol{K}^\alpha(t') \,,
	\label{eq:linear_trajectories}
\end{equation}
with the retarded Greens function $\boldsymbol{\mathcal{G}}^{\mathrm{R}\alpha}$, the one-particle components of which are
\begin{equation}
    \mathcal{G}^{\mathrm{R}\alpha}(t,t') \coloneqq
	\begin{pmatrix}
		g^\alpha_{qq}(t,t') \, \mathbbm{1}_3 \quad & g^\alpha_{qp}(t,t') \, \mathbbm{1}_3 \\
		g^\alpha_{pq}(t,t') \, \mathbbm{1}_3 \quad & g^\alpha_{pp}(t,t') \, \mathbbm{1}_3
	\end{pmatrix} \,.
	\label{eq:greens_function}
\end{equation}
Note that each component is proportional to $\Theta(t-t')$, ensuring the causal structure of the propagator. The interacting part of the action can be expressed in terms of the two collective fields,
\begin{equation}
    S_\mathrm{I}[\vec{\boldsymbol{x}},\vec{\boldsymbol{\chi}}] = - \int \mathrm{d}1\ \vec{\Phi}_B(1) \, \underline{v}(1) \, \vec{\Phi}_n(-1) \,,
    \label{Interacting part of microscopic action}
\end{equation}
where we defined the shorthand notation $\mathrm{d}m\coloneqq\frac{\mathrm{d}^3k_m}{(2\pi)^3} \, \mathrm{d}t_m$ as well as the potential matrix $\underline{v}$ whose components are the pair potentials $v^{\alpha \gamma}$ introduced in \cref{eq:total_interaction_potential},
\begin{equation}
    \underline{v} =
        \begin{pmatrix}
            v^{\ba\ba} & v^{\ba\dm} \\
            v^{\dm\ba} & v^{\dm\dm}
        \end{pmatrix} \,.
    \label{eq:potential_matrix}
\end{equation}
The interaction operator $\hat{S}_\mathrm{I}$ is defined by replacing the collective fields in \cref{Interacting part of microscopic action} with their corresponding operators.

By expanding \cref{free and interacting} in orders of the interaction operator $\hat{S}_\mathrm{I}$, the collective-field cumulants can be computed perturbatively in orders of the interaction potentials. Each order only involves the computation of a finite number of free cumulants, which are exactly known. Details on their computation are given in \cite{fabis_kinetic_2018}. In this paper, we only require the free 2-point cumulants expanded to lowest order in the initial power spectra, as they are the only ones appearing in the computation of the linear power spectrum. They read
\begin{align}
    G^{\alpha \gamma \, (0)}_{BB}(1,2) &= 0 \,,
    \label{2-pt Cumulants 1} \\
    G^{\alpha \gamma \, (0)}_{nB}(1,2) = G^{\gamma \alpha \, (0)}_{Bn}(2,1) &\approx -\im (2\pi)^3 \updelta_\textsc{d}(\vec{k}_1+\vec{k}_2) \, k_1^2 \, \updelta^{\alpha \gamma} \, \bar{n}^\alpha \, g^\alpha_{qp}(t_1,t_2) \,,
    \label{2-pt Cumulants 2} \\
    G^{\alpha \gamma \, (0)}_{nn}(1,2) &\approx (2\pi)^3 \updelta_\textsc{d}(\vec{k}_1+\vec{k}_2) \, \bar{n}^\alpha \bar{n}^\gamma \, \tilde{P}^{\alpha \gamma}(1,2) \,,
    \label{2-pt Cumulants 3}
\end{align}
with the mean number densities of baryonic and dark matter $\bar{n}^\alpha$ as well as
\begin{equation}
    \tilde{P}^{\alpha\gamma}(1,2) \coloneqq P^{\alpha\gamma\,\mathrm{(i)}}_{\delta\delta}(k_1) + \bigl[g^\alpha_{qp}(t_1,0) + g^\gamma_{qp}(t_2, 0)\bigr] \, P^{\alpha\gamma\,\mathrm{(i)}}_{\delta\theta}(k_1) + g^\alpha_{qp}(t_1,0) \, g^\gamma_{qp}(t_2, 0) \, P^{\alpha\gamma\,\mathrm{(i)}}_{\theta\theta}(k_1) \,.
    \label{eq:free_power_spectrum}
\end{equation}

\subsection{Resummed KFT}
As previously introduced for a mixture of baryonic and dark matter in \cite{geiss_resummed_2021}, we will adopt the resummed formulation of KFT (RKFT) developed in \cite{lilow_resummed_2019}. In RKFT the generating functional is reformulated as a path integral over the macroscopic fields of interest, like the density, instead of the underlying microscopic fields $\vec{\boldsymbol{x}}$ and $\vec{\boldsymbol{\chi}}$. This is possible without losing any information on the underlying particle dynamics because the non-interacting dynamics can be solved exactly while the interaction operator \cref{Interacting part of microscopic action} only implicitly depends on the microscopic fields via the collective density and response fields. The reformulation leads to a perturbative expansion in orders of (phase space) density correlation functions, which resums the standard KFT expansion \cref{free and interacting} in orders of the interaction operator.\footnote{Note that unlike in \cite{lilow_resummed_2019} we will not use the full phase space density, as for our purposes the particle number density $n$ is sufficient. Otherwise the procedure is analogous to \cite{lilow_resummed_2019}.} It was found in \cite{geiss_resummed_2019} that this resummation is necessary for a consistent perturbative treatment of mesoscopic particle dynamics.

Mathematically, the reformulation of the generating functional is achieved by introducing an additional path integral over a macroscopic density field $\vec{n} = (n^\ba,n^\dm)$ with a Dirac delta distribution to ensure that $\vec{n}$ describes exactly the same information as the explicitly $\vec{\boldsymbol{x}}$-dependent collective density field $\vec{\Phi}_n$,
\begin{equation}
    \begin{split}
        Z[\vec{\boldsymbol{J}},\vec{\boldsymbol{K}}] &= \int \! \mathcal{D}\vec{n} \; \updelta_\textsc{d}\bigl[\hat{\vec{\Phi}}_n - \vec{n}\bigr] \, Z[\vec{\boldsymbol{J}},\vec{\boldsymbol{K}}] \\
        &= \int \! \mathcal{D}\vec{n} \! \int \! \mathcal{D}\vec{\beta} \, \exp\biggl\{ \im \!\! \int \! \mathrm{d}1 \vec{\beta}(-1) \cdot \bigl[\hat{\vec{\Phi}}_n(1) - \vec{n}(1)\bigr]\biggr\} \, Z[\vec{\boldsymbol{J}},\vec{\boldsymbol{K}}] \,.
    \end{split}
\end{equation}
In the second line we expressed the delta distribution in terms of a path integral over the auxiliary macroscopic field $\vec{\beta}$. Afterwards, any appearance of $\vec{\Phi}_n$ in $Z[\vec{\boldsymbol{J}},\vec{\boldsymbol{K}}]$ can be replaced by $\vec{n}$, which allows us to perform the microscopic $\vec{\boldsymbol{x}}$ and $\vec{\boldsymbol{\chi}}$ integrals in \cref{free and interacting}. Proceeding as in \cite{lilow_resummed_2019}, we then obtain the macroscopic generating functional
\begin{equation}
    Z_\phi[M] \coloneqq \int \! \mathcal{D}\phi \, \exp\biggl\{ \im S_\Delta[\phi] + \im S_\mathcal{V}[\phi] + \int \! \mathrm{d}1 \, M^\top\!(1) \; \phi(-1) \biggr\} \,,
    \label{RKFT Z}
\end{equation}
where we combined the macroscopic fields into $\phi \coloneqq (\vec{n},\vec{\beta})$ and introduced a new associated macroscopic source field $M \coloneqq (\vec{M}_n,\vec{M}_\beta)$. The terms $S_\Delta[\phi]$ and $S_\mathcal{V}[\phi]$ correspond to the propagator and the vertex terms, respectively, adopting the standard nomenclature of statistical and quantum field theory. The propagator term of the action collects all contributions to the action quadratic in $\phi$,
\begin{equation}
    \im S_\Delta[\phi] \coloneqq -\frac{1}{2} \int \! \mathrm{d}1 \! \int \! \mathrm{d}2 \; \phi^\top\!(-1) \; \Delta^{-1}(1,2) \; \phi(-2) \,.
\end{equation}
We can express the inverse propagator $\Delta^{-1}$ in terms of the free collective-field two-point cumulants,
\begin{equation}
    \Delta^{-1}(1,2) = 
    \begin{pmatrix}
       0 &\; \im \, \mathcal{I}(1,2) \mathbbm{1}_2 - \underline{v}^\top(1) \, G^{(0)}_{\vec{B} \vec{n}} (1,2) \\[0.5em]
        \im \, \mathcal{I}(1,2) \mathbbm{1}_2 - G^{(0)}_{\vec{n} \vec{B}} (1,2) \, \underline{v}(2) & G^{(0)}_{\vec{n} \vec{n}} (1,2) \\
    \end{pmatrix} \,,
    \label{inverse propagator}
\end{equation}
where we have introduced a two point identity function
\begin{equation}
    \mathcal{I}(1,2)\coloneqq (2\pi)^3 \, \updelta_\textsc{d}(\vec{k}_1+\vec{k}_2) \, \updelta_\textsc{d}(t_1-t_2) \,,
\end{equation}
and $\mathbbm{1}_n$ denotes the $n \times n$ identity matrix. The vertex term $S_\mathcal{V}[\phi]$ contains all non-quadratic contributions in $\phi$, which are proportional to the different free collective-field $n$-point cumulants with $n \neq 2$. The exact expressions for these can be found in \cite{geiss_resummed_2021}. We will not need them here since we are only interested in the linear power spectrum, which is fully determined by the macroscopic propagator \cite{lilow_resummed_2019}.

The macroscopic-field cumulants are obtained as functional derivatives of the logarithm of the macroscopic generating functional with respect to the macroscopic source field,
\begin{equation}
	G^{\alpha_1 \dotsb \alpha_{l_n} \, \gamma_1 \dotsb \gamma_{l_\beta}}_{n \dotsb n \, \beta \dotsb \beta}(1,\dotsc,l_n,1',\dotsc,l_\beta') = \prod_{u=1}^{l_n} \biggl(\frac{\updelta}{\im \updelta M_n^{\alpha_u}(u)}\biggr) \prod_{r=1}^{l_\beta} \biggl(\frac{\updelta}{\im \updelta M_\beta^{\gamma_r}(r')}\biggr) \ln Z[M]\bigg|_{M=0} \!\! .
	\label{Macroscopic-field cumulants}
\end{equation}
To compute them perturbatively, we expand the generating functional \cref{RKFT Z} in orders of the vertices $S_\mathcal{V}$. A systematic representation of this RKFT perturbation theory in terms of Feynman diagrams was developed in \cite{lilow_resummed_2019}. For the linearly evolved power spectrum, however, we only require the leading-order (tree-level) result of the 2-point density cumulant, which is given by the density-density component of the propagator,
\begin{equation}
    P_{\delta\delta}^{\alpha \gamma \, \text{(tree)}}(k_1,t_1) = \frac{1}{\bar{n}^\alpha \, \bar{n}^\gamma} \int \! \frac{\mathrm{d}^3 k_2}{(2\pi)^3} \int \! \mathrm{d} t_2 \, \updelta_\textsc{d}(t_1-t_2) \, \Delta^\mathrm{\alpha \gamma}_{nn}(1,2) \,.
    \label{eq:individual_tree-level_power_spectra}
\end{equation}
To obtain $\Delta^\mathrm{\alpha \gamma}_{nn}$, the combined matrix and functional inverse of \cref{inverse propagator} needs to be computed, for which we must solve the equation
\begin{equation}
    \int \! \mathrm{d}3 \, \Delta(1,3) \, \Delta^{-1}(-3,2) = \mathcal{I}(1,2) \, \mathbbm{1}_4 \,.
\end{equation}
As described in detail in \cite{lilow_resummed_2019,geiss_resummed_2021}, discretizing the time coordinate into small steps simplifies this into a linear triangular matrix equation, which can be solved numerically inexpensively via forward substitution.

\section{Application to a tightly coupled baryon-photon fluid}
\label{sec:application}
In our previous work \cite{geiss_resummed_2021}, we demonstrated how the two-particle type KFT formulation can be used to describe the joint evolution of dark and baryonic matter in the matter-dominated epoch. The baryons were implemented as an effective mesoscopic particle species describing the collective behavior of baryonic matter in the thermodynamic limit. The need for this effective description of baryons arose from the fact that the frequent small-range interactions of baryonic gas particles cannot be accurately described in low orders of the perturbative expansion. By constructing mesoscopic particles, we average out the particles' microscopic interactions, only keeping their large-scale collective effects in the form of a pressure term.

\begin{figure} 
\begin{center}
\includegraphics[width = 0.8\textwidth]{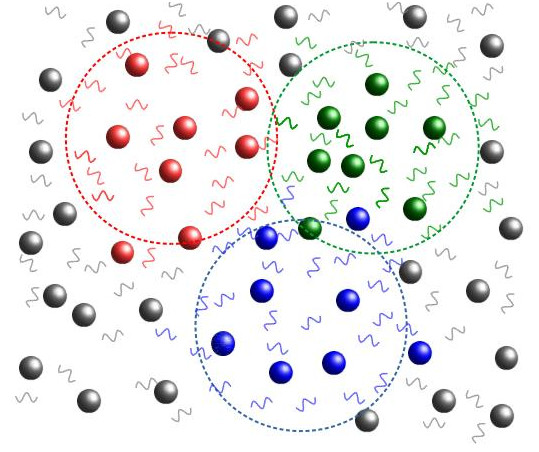}
\caption{Conceptual illustration of mesoscopic particles for a baryon-photon fluid. Baryons are shown as spheres, photons as wiggly lines. The color of each microscopic particle indicates to what mesoscopic particle it belongs. Grey indicates that these particles belong to other mesoscopic particles not emphasized here. \label{BAOmesoscopic}}
\end{center}
\end{figure}

Before hydrogen recombination at a redshift of $z \approx 1100$, photons had a profound influence on the formation of baryonic structures, since the two particle species were tightly coupled. As of now it is not possible to treat photons as a separate particle species within the framework of KFT, since KFT is not (yet) able to incorporate relativistic particles. In the tight-coupling regime we can nevertheless approximate the contribution of photons to the formation of structures, by modifying the mesoscopic particle formalism such that each mesoscopic particle contains both photons and baryons, as illustrated in \cref{BAOmesoscopic}. Due to their extremely high abundance relative to the baryons ($\sim 10^9$ photons per baryon), the pressure of this mesoscopic particle is dominated by its photon content. The mass density of the mesoscopic particles and therefore their gravitational interaction strength, however, has sizeable contributions from both photons and baryons.\footnote{For simplicity, we use the term mass density to refer to the energy density of the photons divided by $c^2$.}

The contribution of photons to the mesoscopic particle mass density decreases with time, since the photons lose energy as the Universe expands. However, in our non-relativistic treatment we cannot consistently describe such an evolution of the mesoscopic particle mass. For that reason we use an approximation in which we average the mass density of photons over the time frame over which we evolve the system. To capture most of the formation history of BAOs while decreasing the error we make in averaging the mass density of photons, we evolve the system from matter-radiation equality at $z_\eq \approx 3400$ up to the decoupling of photons from baryonic matter and thus the end of the tight-coupling regime at $z_\mathrm{dec} \approx 1100$.

Beginning the evolution only at matter-radiation equality changes the resulting sound horizon and therefore the locations of the BAOs. The sound horizon in our approximation is given by
\begin{equation}
    r_{s,\mathrm{approx}} = \int_{t_\eq}^{t_\mathrm{dec}} \frac{\mathrm{d}t}{a(t)} \, c_s(t)
    \label{eq:sound_horizon}
\end{equation}
with $t_\mathrm{dec}$ being the decoupling time, and $t_\eq$ the time of matter-radiation equality, $a(t)$ the scale factor and $c_s(t)$ the speed of sound of the baryon-photon fluid,
\begin{equation}
    c_s(t) = \frac{c}{\sqrt{3 \, \bigl(1+\frac{3 \bar{\rho}^\bary(t)}{4 \bar{\rho}^\rad(t)}\bigr)}} \,.
    \label{speed of sound}
\end{equation}
Here, $c$ is the speed of light, and $\bar{\rho}^\bary$ and $\bar{\rho}^\rad$ are the actual time-dependent mass densities of baryons and photons (radiation), respectively.\footnote{We used the capitalised superscript ``B" for the baryon density to distinguish it from the combined density of the baryon-photon fluid denoted by ``b".} Assuming a Planck-18 cosmology \citep{aghanim_planck_2020-1}, we obtain a value of $r_{s,\mathrm{approx}} = 83 \, \mathrm{Mpc}$. The full sound horizon, obtained by integrating in \cref{eq:sound_horizon} from $t=0$, is $r_s = 136 \, \mathrm{Mpc}$. Therefore, we expect the oscillations in our approximation to be shifted by a significant amount compared to the ones obtained with a Boltzmann solver. For this qualitative demonstration of baryon-photon interactions in KFT, however, this is acceptable. A more accurate description of the BAO positions could be achieved with future developments towards the incorporation of relativistic particles in KFT, allowing us to treat the evolution of structures during the radiation-dominated era.

\subsection{Micro- and mesoscopic particle dynamics}
While the microscopic dark matter particles follow Hamiltonian dynamics, the dynamics of the baryon-photon fluid is described by the hydrodynamic Euler equations. To obtain the equations of motion of the effective mesoscopic particles, the Euler equations need to be projected onto the contributions from individual particles, similarly to the numerical method of Smoothed Particle Hydrodynamics \citep{gingold_smoothed_1977,lucy_numerical_1977}. Following the derivation of this in \cite{geiss_resummed_2019}, and adapting it to the convenient choice of time coordinate $\eta \coloneqq \ln (a/a_\eq)$ discussed in \cref{sec:particle_EOM}, we obtain the expressions for the retarded Green's function \cref{eq:greens_function} and the potential matrix \cref{eq:potential_matrix}, which are needed to calculate the power spectrum.

The components of the Green's function are
\begin{align}
    g^\alpha_{qq}(\eta, \eta') = g^\alpha_{pp}(\eta, \eta') &= \uptheta(\eta - \eta') \,, \\
    g^\alpha_{qp}(\eta, \eta') &= \uptheta(\eta - \eta') \, \int_{\eta'}^\eta \frac{\dif \bar{\eta}}{g(\bar{\eta})} \,, \\
    g^\alpha_{pq}(\eta, \eta') &= 0 \,,
    \label{position momentum propagator}
\end{align}
with the scale function
\begin{equation}
    g(\eta) \coloneqq \ex^{2 \eta} \, \frac{H(\eta)}{H_\eq} \,.
    \label{scale function}
\end{equation}
Here, $H$ denotes the Hubble function and $H_\eq$ its value at the time of matter-radiation equality. During the radiation-dominated epoch considered here, the scale function is
\begin{equation}
    g(\eta) = \ex^{2\eta} \, \sqrt{(\Omega^\dm_\eq + \Omega^\bary_\eq) \, \ex^{-3\eta} + \Omega^\rad_\eq \, \ex^{-4\eta}} \,,
\end{equation}
where $\Omega^\dm_\eq$, $\Omega^\bary_\eq$ and $\Omega^\rad_\eq$ are the dimensionless density parameters of dark matter, baryons and photons (radiation) at matter-radiation equality, respectively. Using $\Omega^\dm_\eq + \Omega^\bary_\eq = \Omega^\rad_\eq = \frac{1}{2}$, this can be further simplified to
\begin{equation}
    g(\eta) = \sqrt{\frac{1+\ex^\eta}{2}} \,.
\end{equation}

The components of the potential matrix are found to be
\begin{align}
	v^{\alpha\dm}(\vec{k},\eta) &= - \frac{\ex^\eta}{g(\eta)} \, \frac{C^\dm_{\mathrm{g}} }{k^2} \,,
	\label{dm potential} \\
	v^{\alpha\ba}(\vec{k},\eta) &= \frac{\ex^\eta}{g(\eta)} \, \bigg[-\frac{C^\ba_{\mathrm{g}} }{k^2} + \delta^{\alpha\ba} \, C_{\mathrm{p}}(\eta) \, \ex^\eta\bigg] \,,
	\label{b potential}
\end{align}
with the gravitational and pressure potential amplitudes
\begin{align}
    C^\gamma_\mathrm{g} &\coloneqq \frac{4 \pi G \, m^\gamma}{a_\eq^3 H_\eq^2} \,,
    \label{c_g} \\
   C_\mathrm{p}(\eta) &\coloneqq \frac{c_s^2(\eta)}{a_\eq^2 H_\eq^2 \, \bar{n}^\ba} \,.
    \label{c_p}
\end{align}
We can further simplify the expression for the gravitational potential amplitude by exploiting that in the thermodynamic limit the individual particle masses $m^\gamma$ are only indirectly connected to physical observables via the mean mass densities $\bar{\rho}^\gamma = m^\gamma \bar{n}^\gamma$. Without loss of generality, we can thus set all particle masses equal, $m^\dm = m^\ba \eqqcolon m$. This mass can then conveniently be expressed in terms of the comoving number and mass densities of dark matter, 
\begin{equation}
    m = \frac{\bar{\rho}^\dm}{\bar{n}^\dm} = \frac{3 \, a_\eq^3 H_\eq^2 \, \Omega^\dm_\eq}{8 \pi G \, \bar{n}^\dm} \,,
\end{equation}
where we expressed $\bar{\rho}^\dm$ via the initial dimensionless dark matter mass density parameter $\Omega^\dm_\eq$. The gravitational potential amplitude \cref{c_g} thus simplifies to
\begin{equation}
    C^\gamma_\mathrm{g} = \frac{3 \, \Omega^\dm_\eq}{2 \, \bar{n}^\dm}
    \label{eq:c_g_simplified}
\end{equation}
for both particle species.

The pressure potential amplitude $C_\mathrm{p}$ in \cref{c_p} depends on the speed of sound $c_s$ defined in \cref{speed of sound}. Replacing $\bar{\rho}^\bary$ and $\bar{\rho}^\rad$ by the corresponding initial dimensionless density parameters results in
\begin{equation}
    c_s^2(\eta) = \frac{c^2}{3 \left(1+\frac{3 \, a(\eta) \, \Omega^\bary_\eq}{4 \, a_\eq \, \Omega^\rad_\eq} \right)} = \frac{c^2}{3 \left(1+\frac{3}{2} \, \ex^\eta \, \Omega^\bary_\eq \right)} \,,
    \label{eq:sound_speed_eta}
\end{equation}
where we again used $\Omega^\rad_\eq = \frac{1}{2}$. $C_\mathrm{p}$ also depends on the mean comoving number density of mesoscopic baryon-photon fluid particles $\bar{n}^\ba$. As described above, we approximate $\bar{n}^\ba$ by its average over the considered time of evolution. For convenience, we express it relative to the mean dark matter number density,
\begin{equation}
    \frac{\bar{n}^\ba}{\bar{n}^\dm} \approx \biggl\langle \frac{\Omega^\bary + \Omega^\rad}{\Omega^\dm} \biggr\rangle_\eta = \frac{\Omega^\bary_\eq}{\Omega^\dm_\eq} + \frac{1}{\eta_\mathrm{dec}} \, \frac{\Omega^\rad_\eq}{\Omega^\dm_\eq} \int_0^{\eta_\mathrm{dec}} \!\!\!\! \dif \eta \, \ex^{-\eta} = \frac{\Omega^\bary_\eq}{\Omega^\dm_\eq} +  \frac{1-\ex^{-\eta_\mathrm{dec}}}{2 \, \eta_\mathrm{dec} \, \Omega^\dm_\eq} \,.
    \label{average mes dens}
\end{equation}
Here, we used that the ratio between baryon and dark matter content does not change, and that by definition $\eta_\eq=0$. Inserting \cref{eq:sound_speed_eta,average mes dens} into \cref{c_p}, we find
\begin{equation}
    C_\mathrm{p}(\eta) = \frac{c^2 \, \Omega^\dm_\eq}{\bar{n}^\dm} \, \biggl[3 \, a_\eq^2 H_\eq^2 \, \biggl( 1 + \frac{3}{2} \,\ex^\eta \, \Omega^\bary_\eq \biggr) \, \biggl( \Omega^\bary_\eq +  \frac{1-\ex^{-\eta_\mathrm{dec}}}{2 \, \eta_\mathrm{dec}} \biggr) \biggr]^{-1} \,.
    \label{eq:c_p_simplified}
\end{equation}

Note that the mean dark matter number density $\bar{n}^\dm$, which both potential amplitudes \cref{eq:c_g_simplified,eq:c_p_simplified} now depend on, cancels out when computing the power spectrum. With our choice of equal micro- and mesoscopic particle masses, only the ratio \cref{average mes dens} between baryon-photon and dark matter densities affects the evolution of the power spectrum.

\subsection{Results}
As the initial density contrast power spectra $P^{\alpha\gamma\,\mathrm{(i)}}_{\delta\delta}$, entering \cref{eq:free_power_spectrum}, we use the cold dark matter Eisenstein-Hu spectrum \cite{eisenstein_baryonic_1998} for both dark matter and the baryon-photon fluid, such that there are initially no BAOs. The spectrum is linearly rescaled to a redshift of 3400, and we adopt the Planck-2018 cosmological parameters \cite{aghanim_planck_2020-1}. The initial spectra involving the momentum divergence, $P^{\alpha\gamma\,\mathrm{(i)}}_{\delta\theta}$ and $P^{\alpha\gamma\,\mathrm{(i)}}_{\theta\theta}$, which also enter \cref{eq:free_power_spectrum}, are directly related to $P^{\alpha\gamma\,\mathrm{(i)}}_{\delta\delta}$ as described in \cref{sec:particle_EOM}.

\begin{figure} 
\begin{center}
\includegraphics[width = 0.8\textwidth]{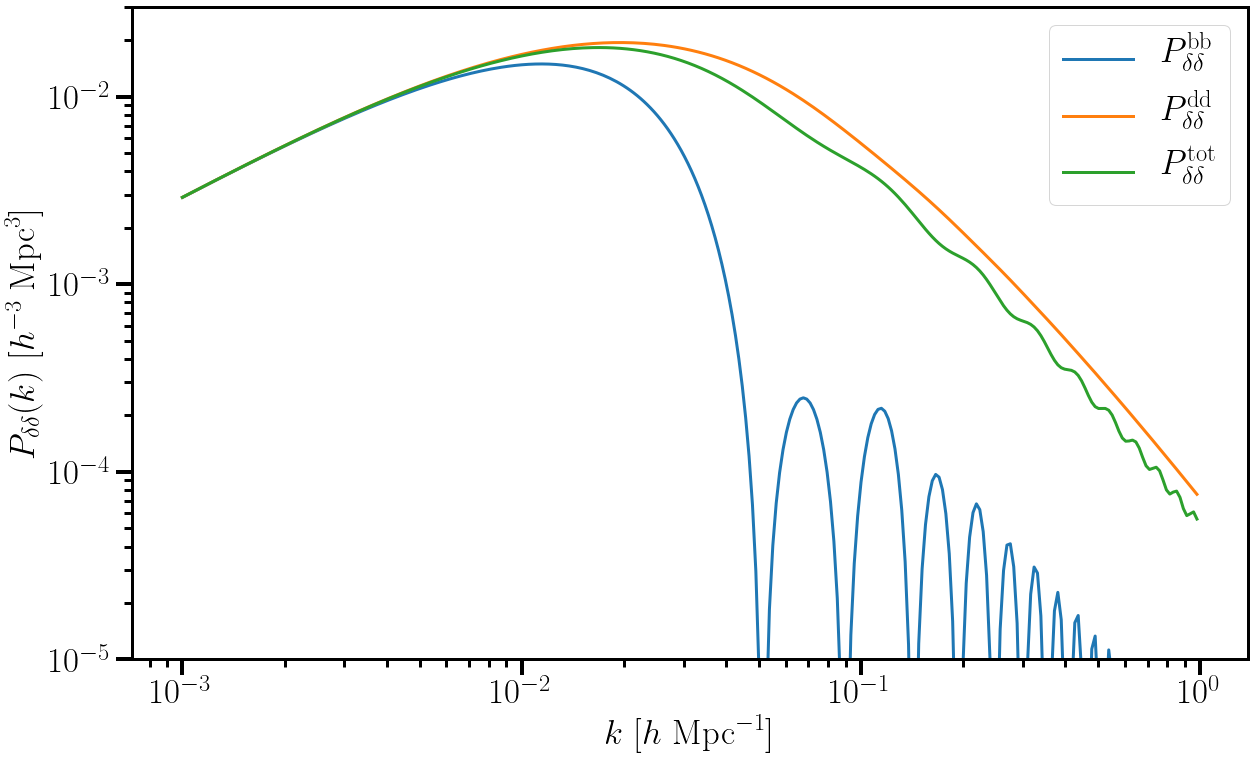}
\caption{Baryon (blue) dark matter (orange) and total matter (green) power spectra obtained in our simplified model by linearly evolving dark matter together with a baryon-photon fluid from matter-radiation equality at $z = 3400$ until recombination at $z = 1100$.
\label{BAO1}}
\end{center}
\end{figure}

Using our results for the interaction potentials \cref{dm potential,b potential,eq:c_g_simplified,eq:c_p_simplified} in the computation of the RKFT propagator and integrating in \cref{eq:individual_tree-level_power_spectra} from the time of matter-radiation equality to photon decoupling ($\Delta \eta = 1.2$), we obtain the dark matter and baryon-photon fluid power spectra plotted in \cref{BAO1}. In addition, we plot the total matter power spectrum of dark and baryonic matter,
\begin{equation}
    P_{\delta\delta}^\mathrm{tot}(k,t) = \frac{(\Omega^\bary_\eq)^2 \, P_{\delta\delta}^{\ba\ba}(k,t) + 2 \, \Omega^\bary_\eq \Omega^\dm_\eq \, P_{\delta\delta}^{\ba\dm}(k,t) + (\Omega^\dm_\eq)^2 \, P_{\delta\delta}^{\dm\dm}(k,t)}{(\Omega^\bary_\eq + \Omega^\dm_\eq)^2} \,,
    \label{total_matter_power_spectrum} 
\end{equation}
using the initial matter density parameters to weigh the individual spectra since their ratio does not change during the evolution.

The main feature in this plot are the oscillations seen in the graph of $P_{\delta\delta}^{\ba\ba}$. They reflect the acoustic oscillations generated by the gravitational attraction of the baryon-photon fluid pushing against the thermal pressure of the photons. We see that odd and even peaks contribute negatively and positively to the total matter power spectrum, respectively. We further note that, relative to the decrease of the total matter power spectrum, the even peaks in $P_{\delta\delta}^{\ba\ba}$ are larger than the odd ones. This phenomenon is caused by baryons contributing to the gravitational interaction but not to the pressure \cite{hu_anisotropies_1995}, such that the ratio of these peaks is related to the baryonic fraction of matter. All these features qualitatively match the expectation for BAOs.

\begin{figure} 
\begin{center}
\includegraphics[width = 0.8\textwidth]{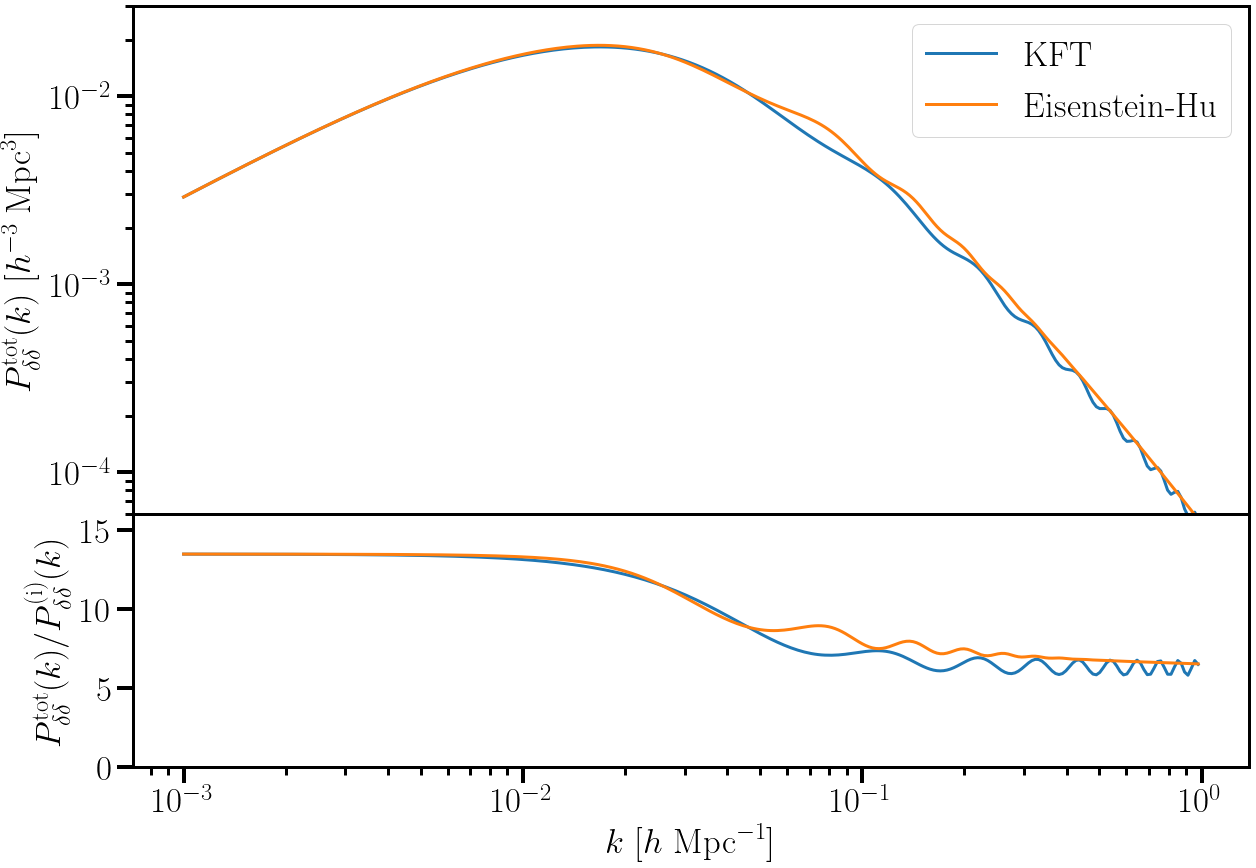}
\caption{\textit{Top:} Comparison between the total matter power spectra at recombination obtained in our simplified model (blue) and using the Eisenstein-Hu transfer function (orange). \textit{Bottom:} The same spectra divided by the initial power spectrum.
\label{comparisonBAO}}
\end{center}
\end{figure}

In \cref{comparisonBAO} we compare the total matter power spectrum obtained from our model to the total matter Eisenstein-Hu power spectrum \cite{eisenstein_baryonic_1998}. Despite the qualitative agreement, we see some notable differences: (i) The first peak in our result is approximately half a wavelength to the right of the Eisenstein-Hu result. As we have already discussed in \cref{sec:application}, this is due to our approximation of initializing the BAO formation only at matter-radiation equality ($z=3400$). In fact, the position of the first peak in our result is consistent with the expected reduced sound horizon $r_{s,\mathrm{approx}} = 83 \, \mathrm{Mpc}$ following from \cref{eq:sound_horizon}. (ii) The suppression of growth due to baryons is larger in our result. This is likely attributed to the approximation we make by setting the mesoscopic mass to a constant value. (iii) At small scales the oscillations in the Eisenstein-Hu spectrum are suppressed due to Silk damping, i.e.~the diffusion of photons at times close to decoupling. Our simplified model of the baryon-photon fluid assumes a tight-coupling regime and thus cannot capture diffusion yet. Further development of the model is needed to address the interactions of photons and baryons at later times.

\section{Discussion \label{discussion}}
We have expanded the analytic mesoscopic particle approach developed to describe baryonic matter in (R)KFT, to capture the physical properties of a baryon-photon fluid in the tight-coupling regime. We then applied this model to describe the co-evolution of dark matter and the baryon-photon fluid between matter-radiation equality and photon decoupling. 

Our results have shown that our simplified analytic model is capable of describing the formation of BAOs in the cosmic matter density power spectrum. The suppression of structure and the appearance of oscillations qualitatively match the features obtained with well-established semi-analytic methods based on the mode decomposition of the Boltzmann equation. Quantitative discrepancies on large scales could be attributed to two approximations we made, namely using matter-radiation equality as our initial time and approximating the time evolution of the energy density of photons by an average value. redBoth of these are consequences of only being able to describe non-relativistic (mesoscopic) particles in KFT so far. On small scales, the suppression of oscillations due to Silk damping was not captured in our model, due to the strict tight-coupling assumption.

One of the most important aspects of the results presented here is that we can use the well-studied structure evolution in the early Universe to properly calibrate the mesoscopic particles in KFT, ensuring that they capture all relevant radiative effects before using them in the investigation of structure evolution at less well-understood eras after decoupling. A crucial aspect of future developments will be the incorporation of photon diffusion, to describe baryon-photon interactions in the post-recombination era. Another direction of future work will be to explore the incorporation of relativistic particles into KFT, necessary for a more accurate description of the pre-recombination era.

\appendix

\section{Particle equations of motion in an extending spacetime}
\label{sec:particle_EOM}
In analogy to the discussion in \cite{bartelmann_trajectories_2015}, the Lagrangian of dark matter and baryon-photon fluid particles in an extending spacetime is given by
\begin{equation}
    L(\boldsymbol{q},\dot{\boldsymbol{q}},t) = \sum_{\alpha=\ba,\dm} \, \sum_{j=1}^{N^\alpha} \, m^\alpha \, \biggl[\frac{a^2}{2} \, (\dot{\vec{q}}^{\,\alpha}_j)^2 - V^\alpha(\vec{q}^{\,\alpha}_j,t)\biggr] \,.
\end{equation}
Here, $a$ is the scale factor, $m^\alpha$ is the mass of a particle of species $\alpha$, and $V^\alpha$ is the overall potential it experiences. The latter splits into gravitational and pressure contributions from individual particles,
\begin{equation}
   V^\alpha(\vec{q}^{\,\alpha}_j,t) = \sum_{\gamma=\ba,\dm}  \, \sum_{l=1}^{N^\gamma} \, v_\mathrm{g}^\gamma(|\vec{q}^{\,\alpha}_j-\vec{q}^{\,\gamma}_l|,t) + \updelta^{\alpha \ba} \sum_{l=1}^{N^\ba} \, v_\mathrm{p}(|\vec{q}^{\,\alpha}_j-\vec{q}^{\,\ba}_l|,t) \,.
   \label{eq:total_potential_split_into_individual_graviational_and_pressure_contributions}
\end{equation}
Both the dark matter and the baryon-photon fluid interact gravitationally, whereas only the baryon-photon fluid contributes to the pressure. As discussed in \cite{geiss_resummed_2019}, the single-particle gravitational and pressure potentials in Fourier space read
\begin{equation}
    v_\mathrm{g}^\gamma(k,t) = - \frac{4 \pi G \, m^\gamma}{a k^2} \,, \quad v_\mathrm{p}(k,t) = \frac{c_s^2}{\bar{n}^\ba} \,,
    \label{eq:individual_graviataional_and_pressure_potentials_before_time_trafo}
\end{equation}
following from the Poisson and Euler equations, respectively, and taking the thermodynamic limit. Here, $c_s$ denotes the speed of sound and $\bar{n}^\ba$ is the mean comoving number density of mesoscopic baryon-photon particles.

We now introduce the time coordinate $\eta = \ln (a/a_\eq)$, with the scale factor at matter-radiation equality, $a_\eq$, and denote derivatives with respect to $\eta$ by a prime. Then the transformed Lagrangian with respect to the new time coordinate is
\begin{equation}
    L_\eta(\boldsymbol{q},\boldsymbol{q}',\eta) = \sum_{\alpha=\ba,\dm} \, \sum_{j=1}^{N^\alpha} \, m^\alpha \, \biggl[\frac{a^2 \, H}{2} \, (\vec{q}^{\,\prime\,\alpha}_j)^2 - \frac{1}{H} \, V^\alpha(\vec{q}^\alpha_j,t)\biggr] \,,
\end{equation}
with the Hubble function $H=\dot{a}/a$. The equations of motion following from the corresponding Hamiltonian read
\begin{equation}
    \vec{q}^{\,\prime\,\alpha}_j = \frac{\vec{p}^\alpha_{\mathrm{can}\,j}}{m^\alpha \, a^2 \, H} \,, \quad \vec{p}^{\,\prime\,\alpha}_{\mathrm{can}\,j} = - \frac{m^\alpha}{H} \, \vec{\nabla}_{q^\alpha_j} \, V^\alpha(\vec{q}^\alpha_j,\eta) \,,
\end{equation}
where $\vec{p}^\alpha_{\mathrm{can}\,j}$ is the canonically conjugate momentum of the $j$-th particle of species $\alpha$. For the purpose of this work, it is more convenient to describe the particles by the rescaled momentum
\begin{equation}
    \vec{p}^\alpha_j = \frac{\vec{p}^\alpha_{\mathrm{can}\,j}}{m^\alpha \, a_\eq^2 \, H_\eq} \,.
\end{equation}
The equations of motion in terms of the new momentum variable are
\begin{equation}
    \vec{q}^{\,\prime\,\alpha}_j = \frac{a_\eq^2 \, H_\eq}{a^2 \, H} \, \vec{p}^\alpha_j \,, \quad \vec{p}^{\,\prime\,\alpha}_j = - \frac{1}{a_\eq^2 \, H_\eq \, H} \, \vec{\nabla}_{q^\alpha_j} \, V^\alpha(\vec{q}^\alpha_j,\eta) \,,
    \label{eq:transformed_equations_of_motion}
\end{equation}
assuming the masses $m^\alpha$ to be time-independent. We can then identify the resulting dark matter and baryon-photon fluid potentials introduced in \cref{eq:total_interaction_potential} by combining equations \cref{eq:total_potential_split_into_individual_graviational_and_pressure_contributions} and \cref{eq:transformed_equations_of_motion},
\begin{equation}
    v^{\alpha\dm}(\vec{k},\eta) = \frac{v_\mathrm{g}^\dm(k,\eta)}{a_\eq^2 \, H_\eq \, H} \,, \quad v^{\alpha\ba}(\vec{k},\eta) = \frac{v_\mathrm{g}^\ba(k,\eta)}{a_\eq^2 \, H_\eq \, H} + \updelta^{\alpha \ba} \, \frac{v_\mathrm{p}(k,\eta)}{a_\eq^2 \, H_\eq \, H} \,.
\end{equation}
Inserting \cref{eq:individual_graviataional_and_pressure_potentials_before_time_trafo} yields the final expressions \cref{dm potential} and \cref{b potential}.

With our choice of coordinates, the negative divergence of the initial momenta is given by
\begin{equation}
    \theta^{\alpha\,(\im)}_j = - \vec{\nabla} \cdot \vec{p}^{\alpha\,(\im)}_j = - \frac{1}{H_\eq} \, \vec{\nabla} \cdot \dot{\vec{q}}^{\alpha\,(\im)}_j = f_\eq \, \delta^{\alpha\,(\im)}_j \,,
\end{equation}
where we used the linearised continuity equation to relate the comoving velocity to the density contrast, with the growth rate 
\begin{equation}
    f = \frac{\dif \ln D_+}{\dif \ln a} \,.
\end{equation}
Accordingly, the initial $\theta \delta$- and $\theta \theta$-power spectra are related to the initial $\delta \delta$-power spectrum via
\begin{equation}
    P^{\alpha\gamma\,\mathrm{(i)}}_{\delta\theta}(k) = f_\eq \, P^{\alpha\gamma\,\mathrm{(i)}}_{\delta\delta}(k) \,, \quad P^{\alpha\gamma\,\mathrm{(i)}}_{\theta\theta}(k) = f_\eq^2 \, P^{\alpha\gamma\,\mathrm{(i)}}_{\delta\delta}(k) \,.
\end{equation}

\bibliography{references_bao,bibliography_RL}

\end{document}